# Distributed Robust State Estimation for Hybrid AC/DC Distribution Systems using Multi-Source Data

Manyun Huang, *Member, IEEE*, Junbo Zhao, *Senior Member, IEEE*, Zhinong Wei, *Member, IEEE*, Marco Pau, *Member, IEEE*, and Guoqiang Sun

*Abstract*—Hybrid AC/DC distribution systems are becoming a popular means to accommodate the increasing penetration of distributed energy resources and flexible loads. This paper proposes a distributed and robust state estimation (DRSE) method for hybrid AC/DC distribution systems using multiple sources of data. In the proposed distributed implementation framework, a unified robust linear state estimation model is derived for each AC and DC regions, where the regions are connected via AC/DC converters and only limited information exchange is needed. To enhance the estimation accuracy of the areas with low measurement coverage, a deep neural network (DNN) is used to extract hidden system statistical information and allow deriving nodal power injections that keep up with the real-time measurement update rate. This provides the way of integrating smart meter data, SCADA measurements and zero injections together for state estimation. Simulations on two hybrid AC/DC distribution systems show that the proposed DRSE has only slight accuracy loss by the linearization formulation but offers robustness of suppressing bad data automatically, as well as benefits of improving computational efficiency.

## I. Introduction

THE increasing penetration of renewable energy sources (RES), as well as the emerging flexible loads, are leading to the transformation of the conventional AC distribution system. The DC-based generations and loads, e.g. photovoltaic (PV) panels and electric vehicles, call for the resurgence of DC grids to improve energy efficiency. An inevitable consequence is the coexistence of the AC and DC grids in the modern distribution system. The resulting hybrid AC/DC distribution systems bring up plenty of new challenges in real-time monitoring, operation, and controlling [1]-[2].

State estimation (SE) is a fundamental tool to extract system states from raw measurements and to validate the operational constraints. This concept was first introduced in power systems by Fred C. Schweppe and one of the representative SE methods is the weighted least square (WLS) [3]. Other robust SE methods against bad data, e.g. the least absolute value (LAV) estimator [4] and the generalized maximum-likelihood estimator [5], have been proposed as well. In contrast to the transmission system with good measurement redundancy, the distribution system has a low coverage of supervisory control and data acquisition (SCADA) measurements. This can lead to unsatisfactory estimation accuracy or even algorithm divergence issue [6]-[7]. A commonly used remedy for that is to build pseudo measurement for distribution system SE (DSSE) via ad hoc modeling approaches [8]-[10]. On the other hand, the advent of smart meters brings very useful information (e.g. load power injection), which allows to enhance the system observability and to monitor the distribution system [11]. Several studies investigated how to incorporate smart meter data in the DSSE. For example, in [12], the smart meter data are aggregated from low-voltage distribution systems and used for medium-voltage DSSE. To further improve the estimation accuracy, [13] proposes to adjust the variance of different smart meters in the DSSE formulation. The time-mismatch between the SCADA measurements (updated every few minutes) and the low timescale smart meter data (updated every half an hour or hourly) also calls for attention [14]-[16]. Note that these issues for the conventional AC distribution system, i.e., the low measurement redundancy and the low refresh rate of smart meters, still exist in the hybrid AC/DC distribution system and will be considered in this work.

Recently, numerous studies were devoted to investigating the power flow [17] and SE [18]-[20] of hybrid AC/DC distribution systems to ensure their secure and economic operation. In the existing literature, the research of SE for hybrid AC/DC distribution systems can be classified into two main categories: centralized and decentralized implementations. The SE solution proposed in this paper belongs to the second category, which estimates the states of AC and DC regions separately. Although the decentralized SE can ensure flexible operating modes and data privacy, it has to iterate alternately to keep the consistency of boundary values at the point of common coupling (PCC) between the AC and DC regions. In [18] and [19], the WLS is used to solve the hybrid AC/DC SE problem via alternated iterations. Such sequential iterations may take a long time to converge due to the nonlinear functions in the SE.

In order to fill the gap, a distributed SE method with the

This work has been supported partially by National Natural Science Foundation of China under Grant U1966205, and by Fundamental Research Funds for the Central Universities under Grant B200201067.

M. Huang, Z. Wei and Q. Sun are with the College of Energy and Electrical Engineering, Hohai University, Nanjing 210098, China (e-mail: hmy_hhu@yeah.net).

J. Zhao is with the Department of Electrical and Computer Engineering, Mississippi State University, Starkville, MS, 39759 USA (e-mail: junbo@ece.msstate.edu).

M. Pau is with the Institute for Automation of Complex Power Systems, RWTH Aachen University (e-mail: mpau@eonerc.rwth-aachen.de).

integration of PMU and SCADA is proposed in [20] and the performance of both the centralized and distributed algorithms are compared and discussed in [21]. Since the deployment of PMUs requires high economic cost and is not prevalent in the distribution level, this paper tries to integrate the smart meter data, zero injections and SCADA together, and proposes a computationally efficient distributed robust SE (DRSE) method for hybrid AC/DC distribution systems.

To capture the fluctuations of states in the hybrid AC/DC distribution system, the proposed DRSE is supposed to be executed every few minutes. In the distributed framework, a unified linear SE model is built for AC and DC regions, and the converter model is derived to ensure the consistency of boundary values. The existing SCADA measurements and smart meter data, which are updated at different timescales, are integrated together to monitor the states accurately with the aid of a DNN-based method. The main contributions of this paper are as follows.

1) The proposed method is executed in a distributed manner and has robustness against noise and bad measurements. It optimally integrates multi-source data with different timescales for hybrid AC/DC distribution system state estimation;

2) A slow timescale smart meters aided DNN model is developed to extract hidden system statistical information and allow deriving nodal power injections that keep up with the real-time measurement update rate;

3) A unified linear SE formulation for multi-regions is derived that achieves higher computational efficiency.

The remainder of this paper is organized as follows. Section II describes the AC/DC system state estimation problem. Then, the distributed AC/DC SE framework is formulated in Section III. Numerical results are given and discussed in Section IV, and finally Section V concludes the paper.

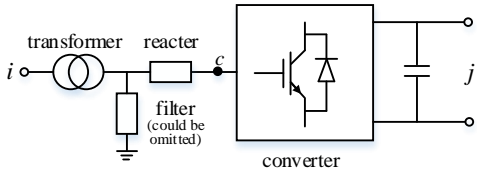

Fig. 1 AC/DC converter model

## II. AC/DC STATE ESTIMATION PROBLEM STATEMENT

Suppose a hybrid AC/DC distribution system with $n$ nodes, i.e., $N = \{1, 2, \ldots, n\}$. These nodes are composed of both AC and DC nodes, denoted as $N_{AC} \subseteq N$ and $N_{DC} \subseteq N$ respectively. Each node can only be of one type, and thereby $N_{AC} \cap N_{DC} = \varnothing$ and $N_{AC} \cup N_{DC} = N$. For the model of AC/DC converters given in Fig. 1, node $i$ belongs to the set of AC nodes (i.e., $i \subseteq N_{AC}$) while node $j$ belongs to the set of DC nodes (i.e., $j \subseteq N_{DC}$). Since the latest converters work without applying the pulse width modulation technology, the filter is not necessary [22]. In this case, only the auxiliary node c is introduced in this paper and assigned as a new AC node, i.e. $N_{AC} \cup c = N_{AC}$.

### A. State Variables

In the SE of hybrid AC/DC distribution systems, the state variables of AC nodes are $x_{AC,i} = \{V_i, \theta_i\}$, where $V_i$ and $\theta_i$ denote the voltage magnitude and angle at the AC node $i$. Analogously, the state variable at the DC node $j$ is the voltage, i.e., $x_{DC,j} = V_j$.

To sum up, the state variables $x$ of a hybrid AC/DC distribution system with node set $N$ can be expressed as:

$$x = \begin{cases} x_{AC,i}, i \in N_{AC} \\ x_{DC,i}, i \in N_{DC} \end{cases} \quad (1)$$

### B. Measurement Functions

In hybrid AC/DC distribution systems, measurements may include power flow, power injection, voltage magnitudes and so on. The measurement model is described as follows:

$$z = h(x) + e \quad (2)$$

where $\mathbf{z} = [z_1, z_2, \ldots, z_m]^T$ is a $m \times 1$ vector, and $e$ is the associated measurement error vector assumed to be white Gaussian noise. $h(\cdot)$ represents the nonlinear relationship between the measurements and states. The measurement functions for a generic branch power and nodal power injections are:

$$P_{AC,ii'} = V_i V_{i'} (G_{ii'} \cos\theta_{ii'} + B_{ii'} \sin\theta_{ii'}), \ \{i,i'\} \in N_{AC} \quad (3)$$

$$Q_{AC,ii'} = V_i V_{i'} (G_{ii'} \sin\theta_{ii'} - B_{ii'} \cos\theta_{ii'}), \ \{i,i'\} \in N_{AC} \quad (4)$$

$$P_{AC,i} = \sum_{i' \in i} P_{AC,ii'} \quad \{i,i'\} \in N_{AC} \quad (5)$$

$$Q_{AC,i} = \sum_{i' \in i} Q_{AC,ii'} \quad \{i,i'\} \in N_{AC} \quad (6)$$

$$P_{DC,jj'} = V_j V_{j'} Y_{jj'}, \ \{j,j'\} \in N_{DC} \quad (7)$$

$$P_{DC,j} = \sum_{j' \in j} P_{DC,jj'} \quad \{j,j'\} \in N_{DC} \quad (8)$$

where $P_{AC,ii'}$ and $Q_{AC,ii'}$ are the active and reactive power flow from the AC node $i$ to the AC node $i'$; $P_{AC,i}$ and $Q_{AC,i}$ are the active and reactive power injection at the AC node $i$; $P_{DC,jj'}$ is the real power flow from the DC node $j$ to the DC node $j'$; $P_{DC,j}$ is the real power injection at the DC node $j$. Parameters G and B are the conductance and susceptance matrices of the AC region while Y is the conductance matrix of the DC region. It is worth noting that the power flow from the converter (omitting the filter) to the AC node $i$ is as follows.

$$P_{AC,ci} - P_{VSC} = 0 \quad (9)$$

$$Q_{AC,ci} - Q_{VSC} = 0 \quad (10)$$

$$P_{VSC} + P_{VSC,loss} = P_{DC,jc} \quad (11)$$

where $P_{AC,ci}$ and $Q_{AC,ci}$ denote the active and reactive power flow from the AC node $c$ to the AC node $i$; $P_{VSC}$ and $Q_{VSC}$ are the active and reactive power outputs of the converter; $P_{VSC,loss}$ is a quadratic loss of the converter [23], and $P_{DC,jc}$ represents the active power flow from DC node $j$ to the converter. In addition, constraints on the associated variables (i.e. $P_{VSC}$, $Q_{VSC}$, $V_{AC,i}$, $V_{DC,j}$) of the AC/DC converter are determined by the control strategy and described in detail in [24].

### C. Objective Functions

The most widely used WLS aims at minimizing the following objective function:

$$J_1 = \sum_{i=1}^{m} w_i (z_i - h_i(x))^2 \quad (12)$$

where $z_i$ is the $i^{\text{th}}$ measurement and $w_i = 1/\sigma_i^2$ is the weight of measurement $z_i$ inversely related to the measurement error variance. Such an approach has high computational efficiency but is vulnerable to bad data. Another representative method, the WLAV, is robust to gross errors and formulated as:

$$J_2 = \sum_{i=1}^{m} w_i |z_i - h_i(x)| = \sum_{i=1}^{m} w_i (u_i + l_i), \quad u_i, l_i \geq 0 \quad (13)$$

where $u_i - l_i = z_i - h_i(x)$; $u_i$ and $l_i$ are non-negative variables. This reformulation transforms the original WLAV optimization problem into a linear programming (LP) problem when all constraints are linear [25].

To this end, the state estimation problem of hybrid AC/DC distribution systems is described as:

$$\begin{aligned}\min_{x} \quad & J_2(x) \\ s.t. \quad & z = h(x) + e, \\ & x = [x_1, x_2, ..., x_i, ...]^T, i \in N\end{aligned} \quad (14)$$

## III. PROPOSED DISTRIBUTED AC/DC STATE ESTIMATION

The general AC/DC SE problem described in (14) can be decoupled into multi-regional SE problems, and the schematic of the proposed DRSE is given in Fig. 2. In the followings, we first formulate the distributed SE algorithm for hybrid AC/DC distribution systems and discuss the mathematical model of AC/DC converters. Then, a unified linear SE is derived for AC and DC regions. Finally, a DNN aided method based on the smart meter data is presented.

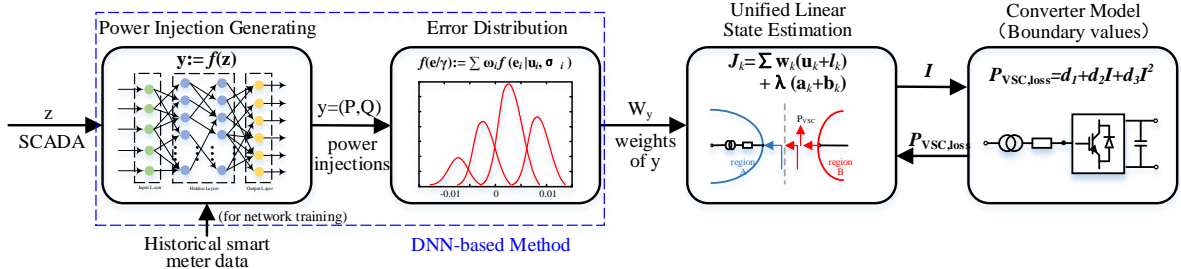

Fig. 2 Schematic of the proposed DRSE

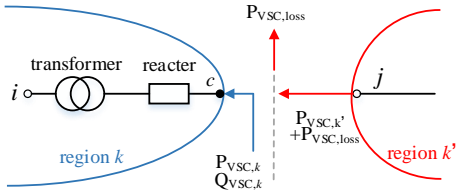

Fig. 3 Power flow at the PCC between region $k$ and $k'$.

### A. Distributed SE Algorithm

Suppose a region $k$ with a node set $N_k$, local state variables $x_k$ and measurements $z_k$, then, the regional SE is described as.:

$$\begin{aligned}\min_{x_k} \quad & J(x_k) \\ s.t. \quad & z_k - h(x_k) = e_k \\ & P_{VSC,k} - P_{VSC,k'} = 0, \quad k' \in D_k\end{aligned} \quad (15)$$

where $k'$ denotes the neighbor region of the region $k$, $D_k$ denotes the set of neighbor regions (the details of the regional partition can be seen in the sample given in Fig. 5). The active power loss has been considered and is shown in Fig. 3, where the AC/DC converter connects node $i$ ($i \subseteq N_{AC}$) with node $j$ ($j \subseteq N_{DC}$) and separates region $k$ from region $k'$. Here, we have two sets of the AC/DC converter output, e.g. $P_{VSC,k}$ and $P_{VSC,k'}$, respectively from region $k$ and $k'$. Assuming the same direction power output of the AC/DC converter in two regions, the boundary values at the PCC should be equal, i.e.,

$$P_{VSC,k} = P_{VSC,k'} \quad (16)$$

$$Q_{VSC,k} = Q_{VSC} \quad (17)$$

Here, the reactive power $Q_{VSC}$ is always decided by the control strategy of the converter (e.g. constant $Q_{VSC}$ or constant $V_{AC,i}$). It can be observed from Fig. 3 that the converter loss, i.e. $P_{VSC,loss}$, is important for ensuring the power balance of AC and DC regions. Specifically, the converter loss heavily depends on the converter current as shown below.

$$P_{VSC,loss} = d_1 + d_2 I_c + d_3 I_c^2 \quad (18)$$

$$I_c = \sqrt{P_{ci}^2 + Q_{ci}^2} / \sqrt{3} V_c \quad (19)$$

where $d_1$, $d_2$ and $d_3$ denote constant loss coefficients; $V_c$, $I_c$, $P_c$, $Q_c$ represent voltage magnitude, current magnitude, active and reactive power injections at the AC node $c$. This quadratic loss formula of the AC/DC converter [22] is employed in the distributed SE framework.

Note that since the active power of the converter is decided by the dependencies between regions, the true value of $P_{VSC,k'}$ in (15) is thus hard to be obtained first. To this end, a relaxed Lagrangian form by using a gradient method is necessary for achieving a fully decoupled distributed algorithm, i.e.,

$$\begin{aligned}\min_{x_k} \quad & J(x_k) + \lambda | P_{VSC,k} - P_{VSC,k'} | \\ s.t. \quad & z_k - h(x_k) = e_k, \quad k' \in D_k\end{aligned} \quad (20)$$

where the Lagrange multiplier $\lambda$ is updated during the iteration.

Finally, the distributed AC/DC SE algorithm based on the WLAV is summarized in Algorithm 1. Here, $P_{loss,k}^l$ denotes the converter loss calculated by the region $k$ in the $l$th iteration; the parameter $\xi$ denotes a positive constant value; the predefined threshold $\tau$ is set to ensure the consistency of boundary values; L denotes the maximum number of iterations and the parameter $R$ denotes the number of AC and DC regions in the hybrid AC/DC distribution system. In this context, only the detailed information of the converter is exchanged between two regions, protecting regional information privacy.

### B. Unified Linear SE for Multi-Regions

In Algorithm 1, each iteration requires the regional SE problem to be solved separately for $R$ regions. To save

```
Algorithm 1: Distributed AC/DC SE
1    Relax the regional SE: J_k = J_2(x_k) + λ|P_{VSC,k} − P_{VSC,k'}|
2    Initialization x_k, λ, ξ, τ, l=0 and L
3    while |P_{VSC,k} − P_{VSC,k'}| > τ and l < L
4        l = l + 1
5        for k=1:D
6            x_k^l := argmin J_k
7            if the region k is an AC region
8                P_{loss,k}^l is obtained by Eq.(18) and (19)
9            end
10       end
11       λ=λ+ξ×|P_{VSC,k} −P_{VSC,k'}|
12   end
13   Output the final solution
```

computing time, the SE process for the AC (or DC) regions is performed in parallel, leading to the following overall computing time for each AC/DC iteration:

$$t_l = \max\{t_k^l, k \in D_{ac}\} + \max\{t_k^l, k \in D_{dc}\} + t_{a,l} \quad (21)$$

where $D_{ac}$ and $D_{dc}$ are the numbers of AC and DC regions respectively, satisfying $D_{ac}+D_{dc}=D$; $t_k^l$ is the calculation time of the regional SE described in step 6 of Algorithm 1; $t_{a,l}$ is the computation time of the algebraic calculation (e.g. steps 8 and 11), and $t_l$ is the computing time of the $l$th iteration. As the algebraic calculation requires little time, the computing time in each iteration mainly depends on the maximum execution time of regional SEs among all regions. Therefore, reducing the calculation time of regional SE in each iteration significantly enhances the computational efficiency of Algorithm 1.

The SE in AC regions is a nonlinear programming problem, and it takes time to converge. To enable an efficient distributed SE algorithm, we extend a linearized power flow model of AC grids [26]-[27] into our work and hence convert the nonlinear SE of AC regions to a linear programming problem. For example, the linear power flow function of a balanced AC network can be expressed as follows:

$$U_i \approx U_j + 2R_{ij}P_{ij} + 2X_{ij}Q_{ij} \quad (22)$$

$$\theta_i \approx \theta_j + X_{ij}P_{ij} - R_{ij}Q_{ij} \quad (23)$$

where $U_i = V_i^2$ is the square of voltage magnitude $V_i$; $R_{ij}$ and $X_{ij}$ denote the resistance and reactance of line $(i, j)$. Note that (22) assumes that the change in voltage associated with losses is negligible, while (23) assumes that voltage magnitude is constant and equal to 1 p.u., as well as the angle difference of two voltage phasors is small enough to satisfy $\sin(\theta_i-\theta_j) \approx \theta_i-\theta_j$. In unbalanced conditions, the voltage magnitude and angle are obtained from three phasors, i.e. $V_{a,b,c}$ and $\theta_{a,b,c}$, and another assumption about the fixed ratio of voltage phasors is made. The accuracy of such approximations has been investigated in [26] and [27]. The validity of the approximation has been demonstrated in our simulation results as well.

With the linearized model, (3) and (4) can be reformulated as below:

$$P_{AC,ij} = \frac{(U_i - U_j)R_{ij} + 2X_{ij}(\theta_i - \theta_j)}{2(R_{ij}^2 + X_{ij}^2)}, \{i,j\} \in N_{AC} \quad (24)$$

$$Q_{AC,ij} = \frac{(U_i - U_j)X_{ij} - 2R_{ij}(\theta_i - \theta_j)}{2(R_{ij}^2 + X_{ij}^2)}, \{i,j\} \in N_{AC} \quad (25)$$

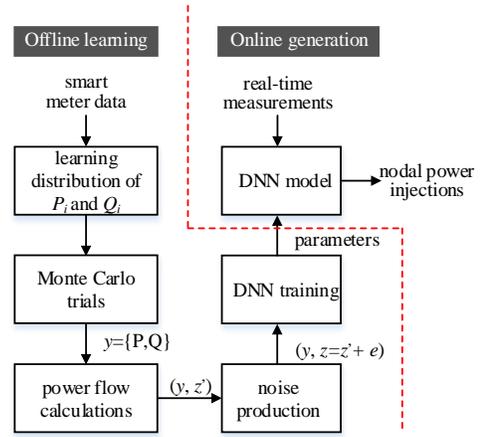

Fig. 4 Schematic of the DNN-based method that learns the distribution of nodal power injections offline and yields nodal power injections online.

In this context, the measurement function in an AC region can be expressed as:

$$z = \mathbf{H}x_{AC}^{new} + e \quad (26)$$

where $z$ denotes the measurement vector in the AC region; $x_{AC}^{new}$ ={$U_i, \theta_i, i \in N_{AC}$} consisting of the squared voltage magnitude and voltage angle represents the state vector in the AC region, and $\mathbf{H}$ is a constant matrix, which depends on the structure and line parameters of the AC region.

As for DC regions, the voltage magnitude can be denoted as $V_i=1-\Delta V_i$, then the power flow in a DC line is:

$$\begin{aligned} P_{dc,ij} &= (1-\Delta V_i)(1-\Delta V_j)Y_{ij} \\ &= (1-\Delta V_i - \Delta V_j)Y_{ij} + \Delta V_i \times \Delta V_j \times Y_{ij} \\ &\approx (1-\Delta V_i - \Delta V_j)Y_{ij} \end{aligned} \quad (27)$$

where the approximation of $P_{dc,ij}$ neglects the high order term, i.e., $\psi(V)=\Delta V_i \times \Delta V_j \times Y_{ij}$. The approximation error associated to $\psi(V)$ will decrease when V approaches 1 (p.u.). In this way, the measurement function in a DC region can be also expressed as a linear measurement function. The effectiveness of such an approximation has been verified in [28].

Formally, the WLAV-based SE for each region as a linear programming problem can be shown as follows:

$$\begin{aligned} \min_{x_k} \quad & J_k = \sum_{i=1}^m w_{k,i}(u_{k,i} + l_{k,i}) + \lambda(a_k + b_k) \\ s.t. \quad & \mathbf{u}_k - \mathbf{l}_k = \mathbf{H}_k x_k + \mathbf{e}_k \\ & a_k - b_k = P_{VSC,k} - P_{VSC,k'} \\ & u_{k,i}, l_{k,i}, a_k, b_k \geq 0, \ k' \in D_k \end{aligned} \quad (28)$$

where $x_k$ and $z_k$ respectively denote the state vector and measurement vector in the region $k$; $\mathbf{H}_k$ is a constant matrix, decided by the structure and line parameters of the region $k$; $\mathbf{e}_k$ denotes the measurement noise vector in the region $k$; $a_k, b_k, u_{k,i}, l_{k,i}$ are non-negative variables associated with the transformation of the objective function. Notice that if the region $k$ is an AC region, the state vector is composed by $x_k$ = {$U_k, \theta_k$}, otherwise, in a DC region, the state vector is $x_k$ = {$V_k$}.

C. DNN-aided method Using Smart Meter Data

As discussed earlier, SCADA measurements, smart meter data and zero injections are together employed in the AC/DC

Table 1 Errors of power injections by the DNN-based method

|  | Active power injection | Reactive power injection |
| --- | --- | --- |
| AAE | 5.29E-04 (p.u.) | 2.31E-04 (p.u.) |
| MAE | 0.0058 (p.u.) | 0.0030 (p.u.) |

SE algorithm to ensure the estimation accuracy. However, they are updated at different timescales and the smart meter data is usually sampled slower than SCADA measurements. To deal with that, this paper proposes a DNN aided method to learn the distribution of nodal power injections from the slow timescale smart meter data (at the offline learning stage) and later on to yield nodal power injections at a fast time scale (at the online generation stage). The rationale of doing this is because once the hidden statistical information from smart meters is effectively extracted, it can be used later to combine with online measurements for estimation accuracy enhancement, avoiding the usage of imprecise pseudo measurements.

The schematic of that is given in Fig. 4, where the left hand of the schematic describes the offline stage. Tracking a large amount of historical smart meter data, the probability distribution of nodal power injections could be approximated by Gaussian mixture modeling (GMM). Then, three submodules, i.e. Monte Carlo trials, power flow calculations and noise production, are subsequently adopted to extract a set of training data ($y$, $z$) from the obtained probability distribution of the nodal power injections. The parameters of the DNN model can be set by the empirical risk minimization. On the basis of the training results, the distribution of power injection errors is approximated by the GMM model to determine the weights of the associated nodal power injections in the following SE. At the online stage, the trained DNN model is used for the regression analysis and to generate nodal power injections associated with the real-time SCADA measurements at the same update rate. Besides, the accuracy of the DNN model could be checked through the deviation of the estimated power injections (calculated by the estimated states) and the outputs y, and the offline training would be triggered once the deviation is larger than a predefined threshold.

## IV. CASE STUDY

The estimation performance of the proposed DRSE is tested on the IEEE 33-bus hybrid AC/DC distribution system [1], while a mesh 106-bus hybrid AC/DC distribution system is applied to demonstrate its scalability [29]. To illustrate the idea of DSSE under a low measurement coverage condition, a few power flow measurements of AC lines are assumed to be available. As for DC regions, we assume that the observability of the DC network is guaranteed by the real-time measurements due to the small size of the DC regions and the lower measurement cost. In the 33-bus hybrid AC/DC distribution system shown in Fig. 5, two types of measurements are assumed: 1) SCADA measurements that are updated every 15 minutes and placed only at some nodes (the main substation node and the converter nodes), four AC lines (line 1-2, 2-19, 3-23 and 6-26), as well as in all DC lines; 2) smart meter data at consumer nodes. With the similar measurement configuration, SCADA measurements in the 106-bus hybrid AC/DC distribution system consist of power flow measurements in 12 AC lines and voltage magnitude measurements in the main substation as well as AC/DC converter nodes, while smart meters are at consumer nodes. In our simulations, the smart meter data is assumed to be sampled hourly at the load/generation nodes and hence updated slower than the SCADA measurements, calling for the generated power injections of the proposed DNN-aided. The additive measurement noises of SCADA measurements are assumed to be Gaussian white noise with 1% uncertainty for the voltage magnitude measurements and 2% for the power flow measurements, respectively. Besides, the smart meter data accuracy is assumed to be 2% [30], and all noises are considered as independent. Note that some communication issues, e.g. the latency, may result in a larger measurement uncertainty of smart meter data and our proposed framework is general to deal with this as well. All simulations were performed on a computer Intel® CoreTM i7-10710U CPU with 16GB of RAM.

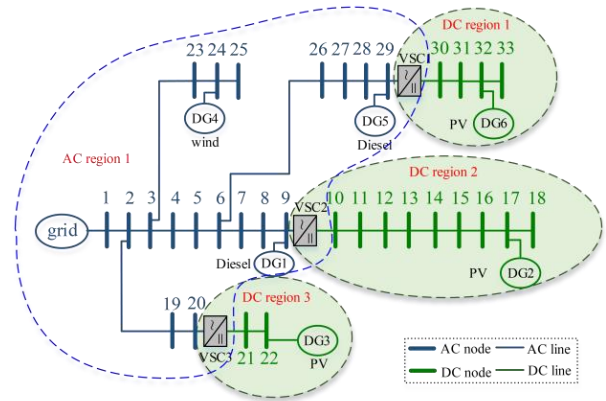

Fig. 5 Structure of the 33-bus hybrid AC/DC distribution system.

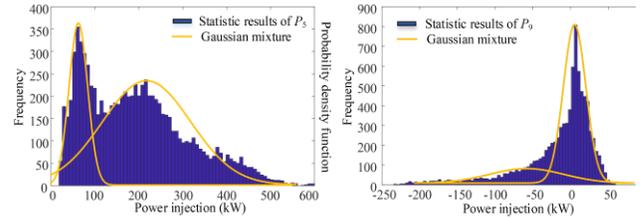

Fig. 6 Distribution learning of the active power injection; left: at node 5; right: at node 9.

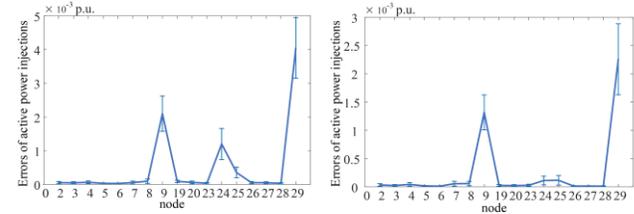

Fig. 7 Errors of power injections at each node; left: active power injections; right: reactive power injections.

Relying on this simulation setup, two alternative algorithms are compared to the proposed DRSE. The first one is the centralized WLS, and the second one is the distributed WLS with the same partition principle as the proposed DRSE shown in Fig. 5. The purpose is to assess the estimation accuracy, the robustness and the computational efficiency of the proposed DRSE. The indices, including the average absolute error (AAE) and maximum absolute error (MAE), are used:

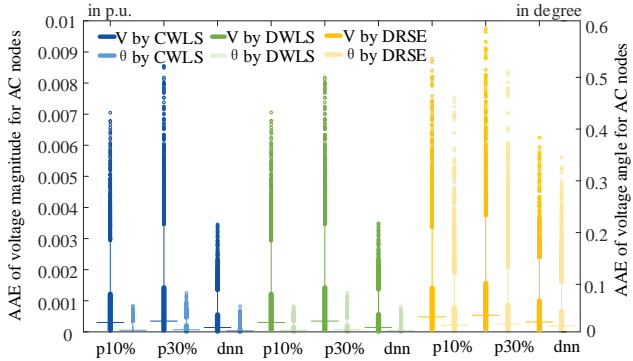

Fig. 8 AAE of AC nodes obtained by three algorithms with different measurement sets.

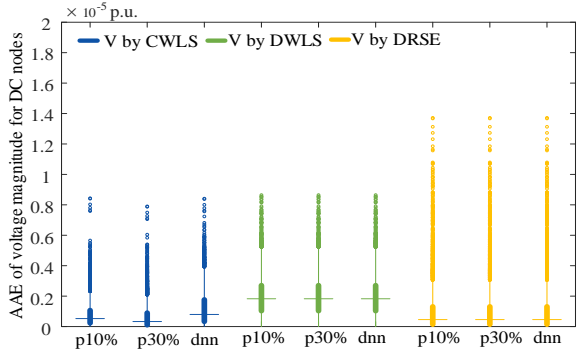

Fig. 9 AAE of DC nodes obtained by three algorithms with different measurement sets.

$$A_x = \frac{1}{n}\sum_{i=1}^{n}|\hat{x}_i - x_i| \qquad (29)$$

$$M_x = \max\{|\hat{x}_i - x_i|, i = 1,...,n\} \qquad (30)$$

where $A_x$ and $M_x$ denote the AAE and MAE of the variable $x$ (i.e., nodal voltage magnitude and voltage angle); $\hat{x}_i$ and $x_i$ represent the estimated values and true values obtained by the power flow calculation.

### A. Validation of DNN-aided Power Injection Generation

Since the load/generation information in [1] only contains one time instant, a set of load/generation profiles collected from Jiangsu Province, covering January 1st of 2013 to April 30th of 2014 are used. Notice that the data set of 2013 is used for distribution learning and DNN training, while that of 2014 is used for testing.

We consider GMM to approximate the distribution of power injections, as shown in Fig. 6. For all nodes, our data analysis allows for the adoption of the 2-component GMM in distribution learning. According to the training performance, the DNN model having four layers with 300 neurons is selected to generate nodal power injections online. The obtained results for the testing errors of nodal power injections are given in Fig. 7, where the active and reactive absolute errors of power injections are respectively smaller than 0.005 (p.u.) and 0.003 (p.u.). It can be found that the errors of power injections at some nodes, i.e. nodes 9, 24 and 29, have large uncertainties due to the fluctuation of distributed energy resources. Again, the distribution of power injection errors was approximated by the GMM model to determine the weights of the associated nodal power injections in the following SE model.

### B. Validation of Estimation Accuracy

To evaluate the estimation accuracy for each method, all three methods are tested under the same measurement conditions. 1) CWLSp: refers to the centralized WLS with pseudo measurements of nodal power injections; 2) CWLSdnn: refers to the centralized WLS with nodal power injections obtained by the proposed DNN-based method; 3) DWLSp: refers to the distributed WLS with pseudo measurements of nodal power injections; 4) DWLSdnn: refers to the distributed WLS with nodal power injections obtained by the proposed DNN-based method; 5) DRSEp: refers to the proposed DRSE with pseudo measurements of nodal power injections; 6) DRSEdnn: refers to the proposed DRSE with nodal power injections obtained by the proposed DNN-based method.

Three SE methods with different accuracy levels of pseudo measurements and the nodal power injections generated by the DNN-based method have been tested. Generally, pseudo measurements are usually obtained by short-term load forecasting with much larger uncertainty than the SCADA measurements. Here, it is assumed that the errors of pseudo measurements follow Gaussian distributions with 10% and 30% uncertainties. Fig. 8 and Fig. 9 show the voltage magnitude and angle estimation results of AC and DC nodes obtained by three algorithms with 1000 Monte Carlo simulations.

It can be observed that the estimation errors of the AC nodes by the three algorithms vary significantly with different measurement uncertainties. The estimation errors under the DNN-generated power injections are smaller than those with pseudo measurements, showing the benefits brought by the DNN model. When using the same measurement set, the estimation errors of voltage angle obtained by our proposed DRSE are larger than those obtained by the CWLS and DWLS due to the linearized process in the AC region (see Fig. 8). This is a compromise between the estimation accuracy and computational efficiency during the linearized process. Furthermore, the estimation errors of voltage angle by the proposed DRSE with the DNN-generated power injections are smaller than 0.4 degree, which is within an acceptable accuracy range. Despite this impact, the estimation errors of voltage magnitudes at the AC and DC nodes by the proposed DRSE are smaller enough when the voltage magnitudes of AC and DC nodes approaching 1 (p.u.). Moreover, it can be seen from Fig. 9 that the estimation errors of the DC nodes under different measurement sets are similar and small. Such estimation performance at the DC nodes is due to the same uncertainties of the DC measurements in the three measuring conditions. To summarize, the proposed DRSE with the DNN-generated power injections could provide reliable estimation results for the hybrid AC/DC distribution system even when the measurement coverage for AC nodes is low avoiding the problem of pseudo measurement definition.

### C. Robustness to Bad Data

Another important property affecting the estimation accuracy is the robustness performance that alleviates the estimation errors brought by bad data. In this regard, three test cases have been considered. 1) Case 1: one active power flow

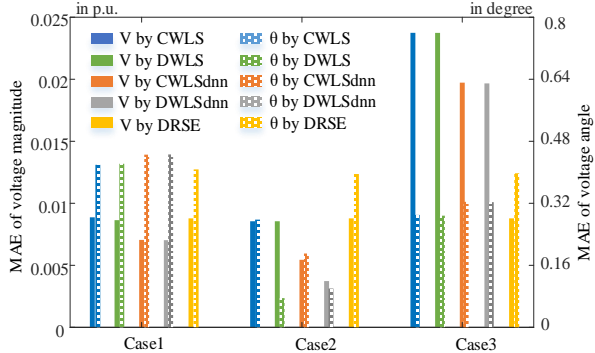

Fig. 10 MAE of AC nodes obtained by the three algorithms in three cases.

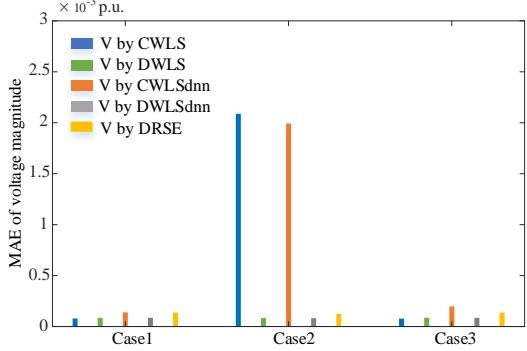

Fig. 11 MAE of DC nodes obtained by the three algorithms in three cases.

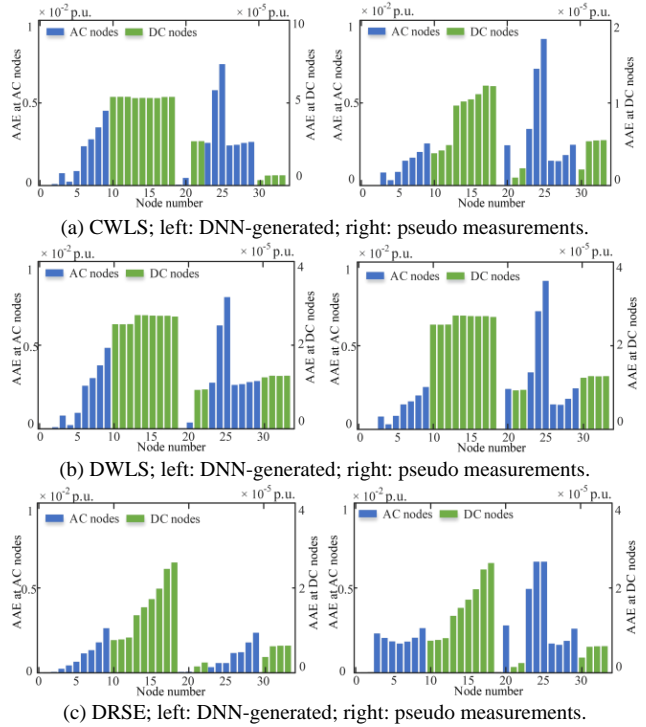

(a) CWLS; left: DNN-generated; right: pseudo measurements.

(b) DWLS; left: DNN-generated; right: pseudo measurements.

(c) DRSE; left: DNN-generated; right: pseudo measurements.

Fig. 12 AAE of each node obtained by three algorithms in Case 3.

measurement of an AC line is doubled. 2) Case 2: one active power output measurement of a converter is doubled. 3) Case 3: a pair of power flow measurement of an AC line becomes the opposite number. For the sake of fairness, the largest normalized residual (NR)-based statistical test has been used in the centralized and decentralized WLS to detect and process bad data. Note that the conventional CWLS and DWLS algorithms with pseudo measurements as well as those with the DNN-generated power injections, namely CWLSdnn and DWLSdnn, are compared with the proposed DRSE. To guarantee the adaptability of the trained DNN model, the erroneous SCADA measurements that are detected and rejected by the NR test in the input are substituted with the values obtained by numerical interpolation. In order to analyze the robustness of the three algorithms, 1000 simulations have been performed for each of the three cases, respectively. Figs. 10 and 11 give the MAE estimates under the three different scenarios.

As expected, the proposed DRSE performs more stable than the other two alternatives in the presence of gross errors. It can be observed that in Cases 1 and 2 the MAE of voltage magnitudes at AC nodes by the proposed DRSE are slightly larger than those of CWLSdnn and DWLSdnn but are comparable to the conventional CWLS and DWLS. Recall that, in normal conditions the linearized model yields slightly larger estimation errors, which could be reduced by the accurate power injections generated by the DNN model. The MAE of DC nodes by the CWLS in each case is the largest one, especially in Case 2. The result indicates that the estimation results of the AC and DC nodes by the centralized CWLS are affected wherever the gross error appears. Instead, the MAE for the DC nodes via two decentralized algorithms remains stable in all cases. Nevertheless, the MAEs for voltage magnitudes by the DWLS are still larger than those by the proposed DRSE in the presence of bad data, especially in Case 3.

To demonstrate the robustness benefits brought by the DNN-generated power injections, pseudo measurements following Gaussian distributions with 30% uncertainties and the DNN-generated power injections are applied for three algorithms in Case 3, and the detailed MAE of each node is shown in Fig. 12. It is observed that the estimation errors by each algorithm with pseudo measurements increase due to the low coverage of real-time measurements and large uncertainty of pseudo measurements. Still, the AAE of each node obtained by the proposed DRSE is smaller than the CWLS and the DWLS. Such estimation performance indicates that the proposed DRSE with the aid of the DNN-generated power injections could provide reliable estimation results in the presence of erroneous measurements.

*D. Computational Efficiency Assessment*

Additional simulations have been conducted on the two hybrid AC/DC distribution systems to assess the computational performance of the proposed DRSE. The computing time includes the execution times of generating the power injections online and that of state estimation. Notice that, multiple AC or DC regions are included in the test system and hence parallel computing is used to decrease the computing time when solving the SE problem for these regions as described in (21). Therefore, two distributed algorithms, i.e., the proposed DRSE and DWLS, are performed 1000 times. The recorded results show that the power injections generated online take only few milliseconds, however, the state estimation takes a longer time and the results are given in Table 2. It could be seen that the computational performance of the proposed DRSE is better than the conventional DWLS. As expect, the execution time of the proposed DRSE is decreased with the linearized model. Moreover, the DWLS needs to be conducted one more time when erroneous measurements are detected by the NR test, while the proposed DRSE is conducted one time only. This allows the proposed DRSE to provide the latest states in real-time for the hybrid AC/DC distribution system even when the SCADA measurements are updated every minute.

Table 2 Execution time of two distributed algorithms

|  | IEEE 33-bus system | Mesh 106-bus system |
|---|---|---|
| DWLS | 138 ms | 203 ms |
| DRSE | 81 ms | 116 ms |

## V. Conclusion

In this paper, a distributed and robust SE method for hybrid AC/DC distribution systems has been developed. The distributed framework allows conducting the regional SE for AC and DC regions separately with only limited information exchange requirement. Moreover, a linearized formulation of the regional SE is derived for AC and DC regions to improve the computational efficiency. A DNN-based method is further proposed to tackle the different timescales of SCADA measurements and smart meter data, avoiding the usage of imprecise pseudo measurements. Simulation results of two hybrid AC/DC distribution systems have demonstrated the effectiveness and scalability of the proposed DRSE. In short, the proposed SE method allows for integrating smart meter data, SCADA measurements and zero injections in the estimation model of hybrid AC/DC distribution systems, protecting regional information privacy, and achieving real-time monitoring.